# Under the same sky with Amanar

Sandra Benítez Herrera[1] and Jorge Rivero González[2] on behalf of the GalileoMobile Team
[1]Instituto de Astrofísica de Canarias, San Cristóbal de La Laguna, Santa Cruz de Tenerife, Spain.
[2]Leiden Observatory, Leiden, the Netherlands.
e-mail: sbenitez@iac.es; rivero@strw.leidenuniv.nl

**The support of the international astronomical community to the 2030 Agenda for Sustainable Development is fundamental to advance the rights and needs of the most vulnerable groups of our global society. Among these groups are the refugees.**

Due to its technological, scientific and cultural dimensions, astronomy is a unique discipline to help achieve the United Nations Sustainable Development Goals [1]. According to the United Nations High Commissioner for Refugees (UNHCR), there are currently nearly 30 million refugees in the world. While there are many (and very necessary) programmes supporting their basic needs, different indicators suggest that the resolution to refugee and internal displacement situations require not only humanitarian interventions, but also development-led actions [2].

One of these initiatives is Amanar: Under the Same Sky, a project designed to support the Sahrawi refugee community by using astronomy to enhance their resilience and engagement in the community, through skill development and self-empowerment activities.

The Sahrawi refugee situation is one of the most protracted in the world, with refugees living in camps near Tindouf, Algeria, since 1975. Access to basic resources is very limited and UN agencies have identified urgent humanitarian needs [2]. On one hand, fresh food, water, and medical and hygiene supplies are scarce in the camps, and Saharawi families depend on international assistance to survive. On the other hand, other identified challenges include the need to ensure continued education for children in the camps. The UNHCR stresses the importance of motivating young people in the camps that are affected by high levels of frustration and limited prospects for their future. In addition, Sahrawi schools face difficulties in terms of quality of teaching as well as lack of educational materials, equipment and infrastructure. Therefore, STEM (science, technology, engineering and mathematics) programmes for capacity building and inspirational activities for the youth as well as teacher training opportunities are essential to support the population within the camps, specifically the educational community and the younger generations [3].

**The Amanar project**

In this context, the GalileoMobile programme [4] established the Amanar project in 2019. GalileoMobile is a volunteer astronomy outreach initiative originated during the UN International Year of Astronomy 2009 [5]. Since its foundation, GalileoMobile has organized actions in 15 countries, sharing astronomy with more than 20,000 people, and has been recognized as a best practice in informal science education by the European Commission [6]. The scope of the programme included activities with underrepresented groups, such as indigenous people in Brazil or communities in territories in conflict, like the separated Greek and Turkish Cypriot populations in Cyprus [7].

Building on these experiences, Amanar aims to inspire Sahrawi youngsters from 8 to 18 through astronomy as well as increase interest in science. The project facilitates the development of scientific skills, such as critical thinking, by performing hands-on activities and sky observations. Moreover, the programme promotes teacher workshops to encourage educators to use astronomy as a didactic tool to contribute to the improvement of the quality of teaching in the region.

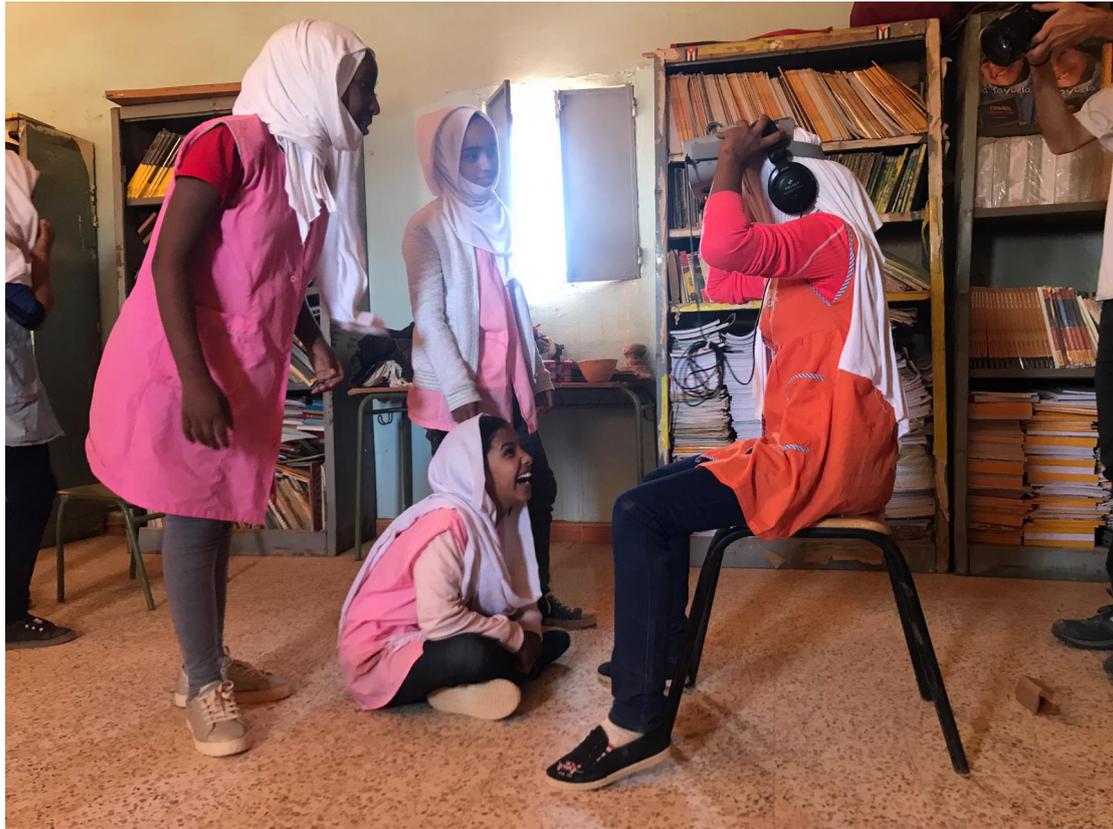

Fig. 1 | Sahrawi girls having fun during an activity using virtual reality (VR). To raise awareness on the struggle of the Sahrawi refugees and allow participants to share their experiences with a global audience, the project is producing an interactive documentary and several VR videos to be released this year. This VR resource is very powerful to generate empathy and is often used in education. Credit: GalileoMobile.

Working in close collaboration with the Instituto de Astrofísica de Canarias (IAC), the Asociación Canaria de Amistad con el Pueblo Saharaui, an organization of solidarity with the Western Sahara from the Canary Islands (Spain), and with the support of the International Astronomical Union Office of Astronomy for Development, the project accomplished two phases in 2019 involving activities both in the Canary Islands and the refugee camps near Tindouf (Algeria).

During the summer, hands-on astronomy activities and visits to professional observatories in the Canary Islands were organized for the Sahrawi children and their host families participating in the Holidays in Peace programme (see below). In October 2019, Amanar organized educational activities and teacher training at the Sahrawi refugee camps and donated kits of educational material (Fig. 1). An ethnographic study on the Sahrawi traditional knowledge of the sky was also conducted. In total, 635 children, 66 teachers and 150 people from the general public participated in the project activities.

**Amanar outcomes**

Even though one should not copy a recipe to the letter when implementing development actions, there are common best practices that can be adopted by similar initiatives. In the following, we discuss the main outcomes of Amanar that we believe may be applicable in other contexts.

- **Local astronomical community as a driver for development**

Separated by only 100 km, the Canary Islands and the Western Sahara share strong historical

and cultural links. Since the start of the conflict, the Canarian civil and political society has repeatedly shown its solidarity with the Sahrawi people. An example is the Holidays in Peace programme in Spain, an initiative to bring Sahrawi children to spend the summer with Spanish host families to escape the harsh conditions of the desert. This programme has been running for over 30 years with the participation of tens of thousands of children.

Another interesting aspect of the region is the excellent conditions for astronomical observations, the Canary Islands being the base of one of the best astronomical observatories in the world.

The Amanar project was conceived taking these factors into account, bringing the large astronomical community residing in the islands to take action and commit to the goal of astronomy for development. Over 20 volunteers from astronomical institutions, such as the IAC, the Gran Telescopio de Canarias and the Cherenkov Telescope Array network, participated in the summer activities in 2019 and those institutions have committed to continue their support of the educational activities with the Sahrawi community in the long term.

- **Learning together under the same sky**

To understand the social, cultural, environmental and political background of the Sahrawi community, the team undertook training sessions about migration and the specifics of the refugee condition, the Sahrawi conflict and the history of the region, given by experts in each area. The last sessions were completed in situ with officials from the UNHCR, the Red Crescent, Sahrawi authorities and local leaders.

This allowed us to know first-hand the challenges the Sahrawi people are facing and to adapt our activities accordingly. For example, a recurrent idea that was brought during the sessions was the feeling of abandonment the population perceives regarding their situation by other countries. The Sahrawi elders also advocate that their culture, based on strong democratic and egalitarian values (for instance, women have equal rights to men and participate in all spheres of society), should be better known in the context of other Arabic nations.

Thus, Amanar performed an educational activity — successfully tested in previous projects — involving the creation of a golden disc, similar to the one carried by the Voyager missions, to share their culture with the rest of the world and the Universe (to see a demonstration of the golden disc activity, please see [here](#)). We observed that for the youth, this appreciation of their culture made them feel proud and recognized.

- **Astronomy as a driver for peace through global citizenship**

A worrying outcome of the stagnation of the Sahrawi conflict is that the younger generations, born and raised at the camps, appear to doubt that diplomatic efforts can resolve the crisis. Every year, more voices are bringing up a military solution for the long-standing situation they have grown up with. Even though the idea of holding a referendum and following a peaceful process towards the liberation of their territory is still the preferred option, the Sahrawi government is losing its trust in the United Nations and is having trouble calming down the youngest.

Therefore, projects like Amanar that use astronomy as a vehicle for encouraging global citizenship focusing on the common aspects that we share as a global society, and endorse the development of critical hope [8], a notion to address unjust systems through meaningful dialogue and empathic responses, constitute a good strategy to be put into practice in territories in conflict. In a turbulent world as the one we live nowadays, this aspect is fundamental for preventing the growth of hate and keeping peace in the long term [9].

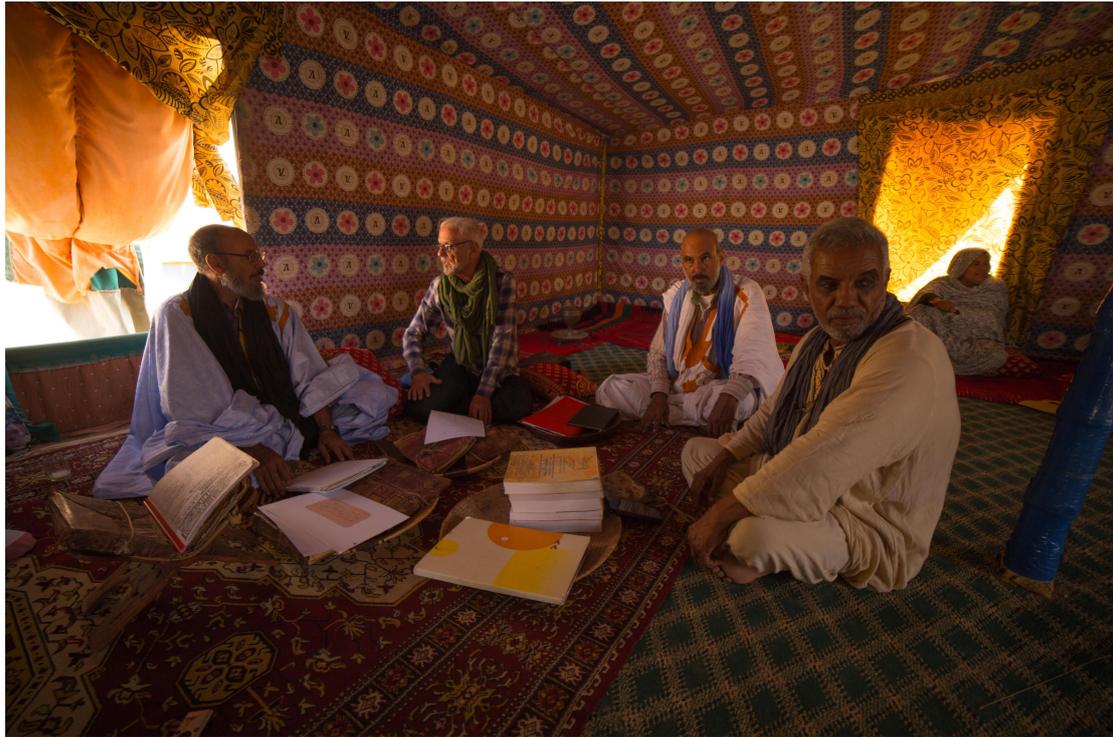

Fig. 2 | Descendants of the Sahrawi philosopher and poet Chej Mohammed el Mami. The Sahrawi knowledge about the sky has passed from one generation to another but it is in danger of disappearing due to the modern ways of life and the impossibility of the Sahrawi to return to their homeland. Credit: GalileoMobile.

- **A sky without borders**

The knowledge of the sky always played a role within the Sahrawi cultural tradition, since the Bedouins used stars and constellations for guidance, time-keeping and to predict changes in climatic conditions. However, most of this wisdom is still transmitted orally and due to the modern ways of life it is in great danger of disappearing.

Hence, Amanar also aimed to support research about the astronomical culture of the Sahrawi people, by interviewing elders and intellectuals (Fig. 2), not considering it as a mere case study, but as a living knowledge and a valuable worldview. In this way, the sky becomes a central axis to explore their collective memory and identity, a territory that has no borders, where they can roam freely, as opposed to the land in which they have no place, no country.

The team has started a collaboration with the Oral History Department from the Sahrawi Ministry of Culture, to support them with the necessary resources (technical equipment, training and fellowships for their personnel) so they can lead this research on their astronomical tradition, use it to reconstruct their past in order to build their future, and engage the youth in the process.

**Leaving no one behind**

As the Amanar team works on an action plan to allow the sustainability of the project, we would like to encourage the international astronomical community to contribute to the awareness of the Sahrawi situation and other similar conflicts around the world. As we reach for the stars, we need to make sure that we don't leave anyone behind.

**Acknowledgements**

Above all, Amanar is a story of solidarity. It would not have been possible without the support of the Canarian Association of Friendship with the Saharaui People and the funds of the International Astronomical Union (IAU) through the Office of Astronomy for Development (OAD) and its centenary celebrations (IAU100), and the Instituto de Astrofísica de Canarias. In addition, we are grateful to our international collaborators: the Cherenkov Telescope Array network, the Gran Telescopio de Canarias, the Virgo collaboration, the Spanish Translation Network of the IAU, the Office Astronomy Outreach, the astronomical association AMNIR and the Superior Council of Scientific Research as well as to all the Canarian local collaborators: Titsa, Cielos de la Palma, CEIP en Arucas, Asociación Canarias de Solidaridad con el Pueblo Saharaui, Fundación Canaria Observatorio de Temisas y Agrupación Astronómica de Gran Canaria.